\documentclass[sigconf,table]{acmart}

\AtBeginDocument{%
  \providecommand\BibTeX{{%
    \normalfont B\kern-0.5em{\scshape i\kern-0.25em b}\kern-0.8em\TeX}}}

\copyrightyear{2022}
\acmYear{2022}
\setcopyright{rightsretained}
\acmConference[CHI '22]{CHI Conference on Human Factors in Computing Systems}{April 29-May 5, 2022}{New Orleans, LA, USA}
\acmBooktitle{CHI Conference on Human Factors in Computing Systems (CHI '22), April 29-May 5, 2022, New Orleans, LA, USA}
\acmDOI{10.1145/3491102.3501951}
\acmISBN{978-1-4503-9157-3/22/04}

\acmSubmissionID{****}

\usepackage{color}
\usepackage{soul}    % highlighting
\usepackage{gensymb} % degree symbol

\usepackage{pifont}% http://ctan.org/pkg/pifont
\usepackage{tikz}
\definecolor{Red}{rgb}{1., 0, 0}
\definecolor{Blue}{rgb}{0, 0, 1.}
\definecolor{Green}{rgb}{0., .6, 0.}
\definecolor{Custom}{rgb}{0.3, .1, 0.2}
\definecolor{Yellow}{rgb}{.9, .7, 0.}
\definecolor{Purple}{rgb}{.9, .1, 0.8}

\graphicspath{{figures/}}

\begin{document}

%%
%% The "title" command has an optional parameter,
%% allowing the author to define a "short title" to be used in page headers.

\title[InfraredTags: Embedding Invisible AR Markers and Barcodes]{InfraredTags: Embedding Invisible AR Markers and Barcodes Using Low-Cost, Infrared-Based 3D Printing and Imaging Tools}

% \title[InfraredTags: 3D Printed Invisible Codes Fabricated from Infrared-Transmitting Filaments]{InfraredTags: 3D Printed Invisible Codes \\ Fabricated from Infrared-Transmitting Filaments and \\ Rapidly Detected by Low-Cost Cameras}

% \title[InfraredTags: Embedding  Invisible, Fast Scanning AR Markers \& Barcodes Into 3D Printed Objects ]{InfraredTags: Embedding  Invisible, Fast Detected \\ AR Markers \& Barcodes Into 3D Printed Objects Using \\ Infrared-Transmitting Filaments}

%\title[Infrared: Embedding Information in Structural Elements and Mechanisms of Laser-Cut Objects]{StructCode: Embedding Information in \\  Structural Elements and Mechanisms of Laser-Cut Objects}
%\title{InfraredTags: 3D Printed, Instant-Read Invisible Codes \\ Using Low-Cost Infrared Cameras}

% \title[InfraredTags: Rapidly Readable Invisible Codes Using Infrared-Transmitting 3D Printing Filaments]{InfraredTags: Rapidly Readable Invisible Codes \\ Using Infrared-Transmitting 3D Printing Filaments \\ and Low-Cost Infrared Cameras}
% Invisible Codes 3D Printed with Infrared Filament
% Using infrared filament to 3D print invisible code
% Fast-Readable
% Invisible Codes 3D Printed with Infrared-Transmitting Filaments that can be Quickly Read Using Low-Cost IR Cameras

\author{Mustafa Doga Dogan}
\affiliation{%
  \institution{MIT CSAIL}
  \city{Cambridge, MA}
  \country{USA}
  }
\email{doga@mit.edu}

\author{Ahmad Taka}
\affiliation{%
  \institution{MIT CSAIL}
  \city{Cambridge, MA}
  \country{USA}
  }
\email{ahmadtak@mit.edu}

\author{Michael Lu}
\affiliation{%
  \institution{MIT CSAIL}
  \city{Cambridge, MA}
  \country{USA}
  }
\email{mlu0708@mit.edu}

\author{Yunyi Zhu}
\affiliation{%
  \institution{MIT CSAIL}
  \city{Cambridge, MA}
  \country{USA}
  }
\email{yunyizhu@mit.edu}

\author{Akshat Kumar}
\affiliation{%
  \institution{MIT CSAIL}
  \city{Cambridge, MA}
  \country{USA}
  }
\email{akshat1k@mit.edu}

\author{Aakar Gupta}
\affiliation{%
  \institution{Facebook Reality Labs}
  %\streetaddress{1 Th{\o}rv{\"a}ld Circle}
  \city{Redmond, WA}
  \country{USA}
  }
\email{aakarg@fb.com}

\author{Stefanie Mueller}
\affiliation{%
  \institution{MIT CSAIL}
  \city{Cambridge, MA}
  \country{USA}
  }
\email{stefanie.mueller@mit.edu}

\renewcommand{\shortauthors}{Dogan et al.}

\begin{abstract}
Existing approaches for embedding unobtrusive tags inside 3D~objects require either complex fabrication or high-cost imaging equipment.
We present InfraredTags, which are 2D markers and barcodes imperceptible to the naked eye that can be 3D printed as part of objects, and detected rapidly by low-cost near-infrared cameras. We achieve this by printing objects from an infrared-transmitting filament, which infrared cameras can see through, and by having air gaps inside for the tag's bits, which appear at a different intensity in the infrared image.

We built a user interface that facilitates the integration of common tags (QR codes, ArUco markers) with the object geometry to make them 3D printable as InfraredTags. We also developed a low-cost infrared imaging module that augments existing mobile devices and decodes tags using our image processing pipeline.
Our evaluation shows that the tags can be detected with little near-infrared illumination (0.2lux) and from distances as far as 250cm. 
We demonstrate how our method enables various applications, such as object tracking and embedding metadata for augmented reality and tangible interactions.

\end{abstract}

%%
%% The code below is generated by the tool at http://dl.acm.org/ccs.cfm.
%% Please copy and paste the code instead of the example below.
%%


%%
%% Keywords. The author(s) should pick words that accurately describe
%% the work being presented. Separate the keywords with commas.

%% A "teaser" image appears between the author and affiliation
%% information and the body of the document, and typically spans the
%% page.
\begin{teaserfigure}
  \centering
  \includegraphics[width=0.95\textwidth]{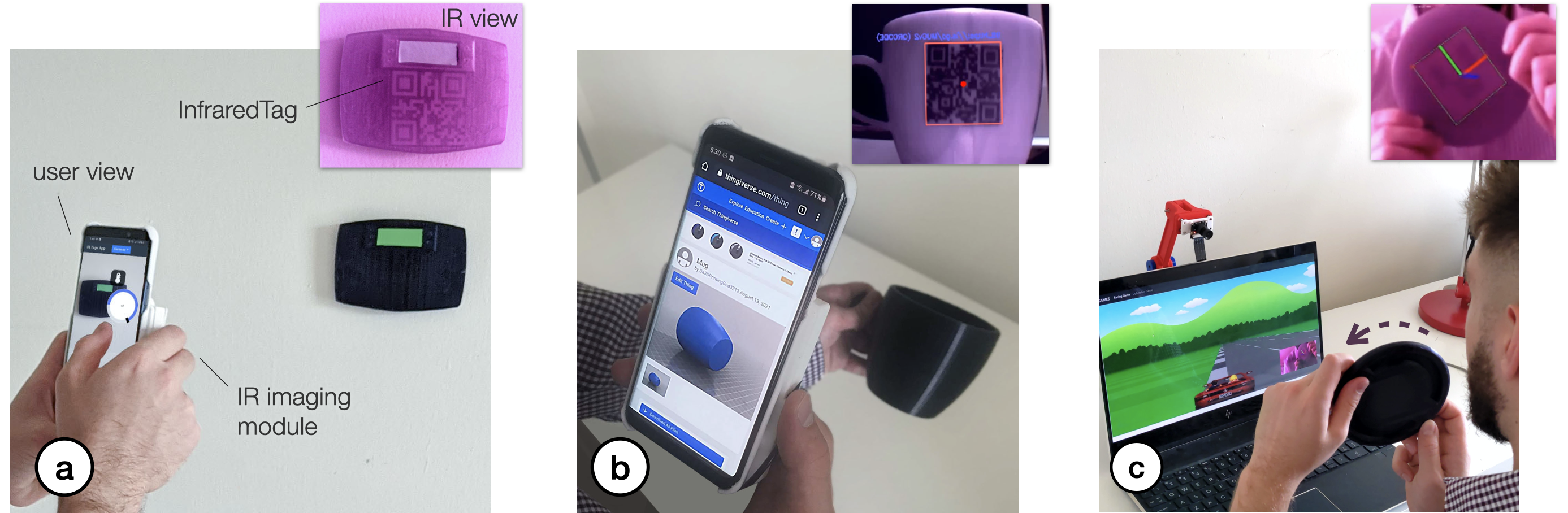}
  \caption{InfraredTags are 2D markers and barcodes embedded unobtrusively into 3D printed objects and can be detected using infrared cameras (top-right images). 
  %InfraredTags are 2D markers and barcodes embedded unobtrusively into objects using infrared-based 3D printing and imaging tools.
  %While (b)~invisible to the naked eye, they can be detected using low-cost infrared cameras.
  This allows real-time applications for (a) identifying and controlling devices in AR interfaces, (b) embedding metadata such as 3D model URLs into objects, and (c) tracking passive objects for tangible interactions.}
  \Description{(a) A person holding a phone towards a thermostat. The phone shows the temperature control menu. (b) A person holding a phone towards a black mug. The phone shows a Web page that has the 3D model of the mug. (c) A man holding a small wheel facing a computer where a driving game is playing. On the top-right corner of each image, we see the infrared view which reveals the embedded 2D tags.}
  \label{fig:teaser}
\end{teaserfigure}

% (a)~Common tags, such as QR codes and ArUco markers, can be embedded as InfraredTags into 3D~printed objects with our user interface.
%   %are embedded (a) underneath the surface of 3D objects.
%   While (b)~invisible to the naked eye, they can be detected using low-cost infrared cameras. (c)~This allows real-time applications for tracking and controlling devices or embedding metadata in objects.

%%
%% This command processes the author and affiliation and title
%% information and builds the first part of the formatted document.
\maketitle

\section{Introduction}

The ability to embed unobtrusive tags in 3D objects while they are being fabricated is of increasing relevance due to its many applications in augmented and virtual reality (AR/VR), packaging, tracking logistics, and robotics.
%Embedding unobtrusive tags in 3D objects is an increasingly important approach for enabling novel interactions with them.

In the last decade, researchers have investigated several ways to insert tags that are imperceptible to the naked eye. One method to accomplish this is to leave air gaps inside the object that represent the bits of a tag. For instance, \textit{AirCode}~\cite{li_aircode_2017} embeds air gaps underneath the surface of 3D printed objects and uses scattering of projected structured light through the material to detect where the air gaps are located. \textit{InfraStructs}~\cite{willis_infrastructs_2013} also embeds air gaps into the object but scans it in 3D using terahertz imaging, which can penetrate better through material than visible light. 

While both of these methods can embed tags inside 3D objects, they require complex hardware setups (e.g., a projector-camera setup as in \textit{AirCode}), expensive equipment (e.g., a terahertz scanner as in \textit{InfraStructs}), and long imaging time on the order of minutes. To address these issues, we propose a new method that combines air gaps inside the 3D printed structure with infrared transmitting filament. This makes the object semitransparent, and the air gaps are detectable when viewed with an infrared camera. Thus, our method only requires a low-cost infrared imaging module, and because the tag is detected from a single frame, scanning can be achieved much faster. 

One method that has used infrared-based 3D printing materials is \textit{LayerCode}~\cite{maia_layercode_2019}, which creates 1D barcodes by printing objects from regular resin and resin mixed with near-infrared dye. Thus, while the printed objects look unmodified to humans, infrared cameras can read the codes. However, this method required a modified SLA printer with two tanks, custom firmware, and custom printing material.
%In contrast, our method uses an off-the-shelf 3D printing filament and works on standard FDM printers. % with the regular slicers. 
In contrast, our method uses more readily available low-cost materials.

In this paper, we present InfraredTags, a method to embed markers and barcodes in the geometry of the object that does not require complex fabrication or high-cost imaging equipment. We accomplish this by using off-the-shelf fused deposition modeling (FDM) 3D printers and a commercially available infrared (IR) transmitting filament~\cite{3dkberlin_pla_2021} for fabrication, and an off-the-shelf near-infrared camera for detection. 
The main geometry of the object is 3D printed using the IR filament, while the tag itself is created by leaving air gaps for the bits. Because the main geometry is semitransparent in the IR region, the near-infrared camera can see through it and capture the air gaps, i.e., the marker, which shows up at a different intensity in the image. The contrast in the image can be further improved by dual-material 3D printing the bits from an infrared-opaque filament instead of leaving them as air gaps. Our method can embed 2D tags, such as QR codes and ArUco markers, and can embed multiple tags within the object, which allows for scanning from multiple angles while tolerating partial occlusion. To be able to detect InfraredTags with conventional smartphones, we added near-infrared imaging functionality by building a compact module that can be attached to existing mobile devices.

To enable users to embed the tags into 3D objects, we created a user interface that allows users to load tags into the editor and position them at the desired location. The editor then projects the tags into the 3D geometry to embed them with the object geometry. After fabrication, when the user is taking a photo with our imaging module, our custom image processing pipeline detects the tag by increasing the contrast to binarize it accurately. This enables new applications for interacting with 3D objects, such as remotely controlling appliances and devices in an augmented reality (AR) environment, as well as using existing passive objects as tangible game controllers.

\vspace{0.2cm}
In summary, our contributions are as follows:
\begin{itemize}
    \item A method for embedding invisible tags into physical objects by 3D printing them on an off-the-shelf FDM 3D printer using an infrared transmitting filament.

    \item A user interface that allows users to embed the tags into the interior geometry of the object.
    
    \item An image processing pipeline for identifying the tags embedded inside 3D prints.
    
    \item A low-cost and compact infrared imaging module that augments existing mobile devices.
    
    \item An evaluation of InfraredTags detection accuracy based on 3D printing and imaging constraints.
    
\end{itemize}

\section{Related Work}

In this section, we first explain how tags have been used in HCI, what kind of approaches have been proposed to make them less obtrusive, and how infrared imaging has been used for different purposes in existing work.

\begin{table*}%[]
\begin{tabular}{@{}c@{}}
\includegraphics{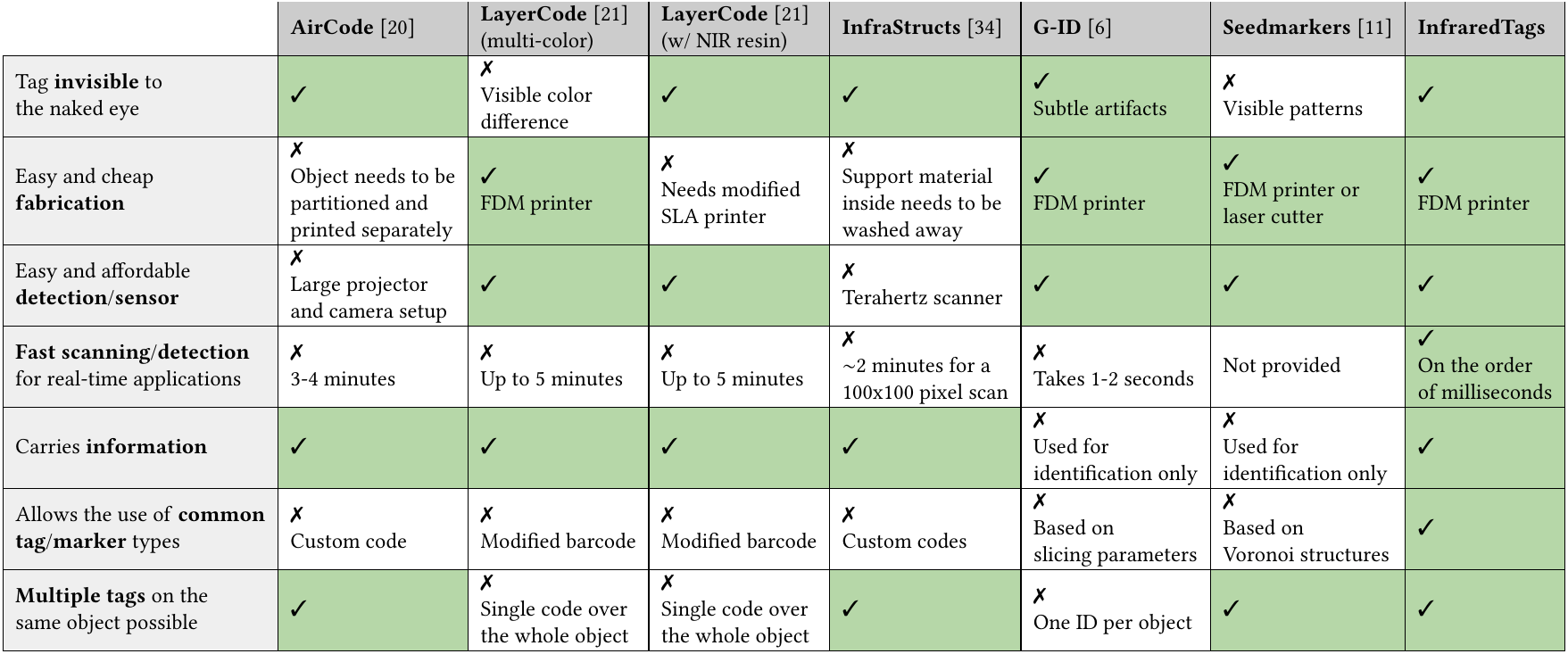}
\end{tabular}
\vspace{0.2cm}
\caption{A comparison of features of different tag embedding methods. Exact values are reported when possible.}
\label{tab:LiteratureComparion}
\vspace*{-10pt}
\end{table*}

\subsection{Use Cases for Tags in HCI}
Tags have been used to mark objects and enable different interactive applications with them.
For instance, \textit{Printed Paper Markers~}\cite{zheng_tangible_2020} use different paper structures that conceal and reveal fiducial markers (i.e., \textit{ArUco}~\cite{romero-ramirez_speeded_2018}) to create physical inputs, such as buttons and sliders.
\textit{DodecaPen}~\cite{wu_dodecapen_2017} can transfer users' handwriting to the digital environment by tracking ArUco markers attached on a passive stylus.
\textit{Cooking with Robots}~\cite{sugiura_cooking_2010} uses detachable markers to label the real-world environment for human-robot collaboration. %to achieve a cooperative tasks for humans and robots.
\textit{Position-Correcting Tools}~\cite{rivers_position-correcting_2012} scan QR code-like markers to precisely position CNC tools while users cut sheets.

Another major use case of markers in HCI is tangible interaction on surfaces. For example, \textit{TUIC}~\cite{yu_tuic_2011} enables such interaction on capacitive multi-touch devices using 2D tags with conductive materials that simulate finger touches.
To build haptic control interfaces, \textit{ForceStamps}~\cite{han_forcestamps_2020} uses 3D printed fiducial markers 
 and \textit{PneuModule}~\cite{han_pneumodule_2020} uses pneumatically-actuated inflatable pin arrays.
\textit{CapCodes}~\cite{gotzelmann_capcodes_2016} and \textit{BYO*}~\cite{gunther_byo_2017}suggests tangibles 3D printed with conductive filaments to enable their identification when touched or moved on a touch display.
\textit{CapacitiveMarker}~\cite{ikeda_capacitivemarker_2015} consists of both a visual marker and a conductive pattern, allowing it to be recognized by both cameras and capacitive displays.
However, all these markers are visible to the human eye which impacts object aesthetics and may reduce the usable area on the object.

\subsection{Making Markers Less Obtrusive}
Researchers have investigated two primary approaches to make markers less obtrusive: making visible tags more \textit{aesthetic}~\cite{qiao_structure-aware_2015, bencina_improved_2005, preston_enabling_2017, getschmann_seedmarkers_2021}, \textit{embedding} tags inside the object~\cite{li_aircode_2017, willis_infrastructs_2013, dogan_g-id_2020}, or having tags that are \textit{invisible} to the human eye on the surface of the object~\cite{maia_layercode_2019, dogan_sensicut_2021}. 

%\textbf{Making them more aesthetic:} ~\cite{preston_enabling_2017}
To make codes more aesthetic, researchers have modified traditional QR codes (\textit{halftoned}) to look more like an aesthetic image (e.g., a photo) while still preserving its detectability~\cite{qiao_structure-aware_2015}.
\textit{ReacTIVision}~\cite{bencina_improved_2005} creates fiducial markers that look like amoeba to create an organic look.
\textit{Seedmarkers}~\cite{getschmann_seedmarkers_2021} are decorative markers manufactured to fit onto an object's desired surface, such as one of the plates of a laser-cut box.

Tags can also be embedded inside 3D objects so that users cannot see them. For example, \textit{AirCode}~\cite{li_aircode_2017} leaves air gaps underneath the object surface to represent the bits of the tags, and uses a camera and projector to decode them. However, this method takes 3-4 minutes to decode the code because it requires to sweep the projection pattern multiple times over the code to reduce noise.  \textit{InfraStructs}~\cite{willis_infrastructs_2013} encodes information inside objects by leaving air gaps inside the object and detecting these by using a Terahertz scanner. However, this scanner needs multiple minutes to capture an image.

Finally, tags can also be embedded unobtrusively on the surface of 3D objects by making them invisible to users. \textit{G-ID}~\cite{dogan_g-id_2020}, for instance, varies the slicing parameters to modify an object's surface texture. Because the changes are subtle, the identifiers are unobtrusive to the user. \textit{LayerCode}~\cite{maia_layercode_2019} makes 1D barcodes on the surface invisible to the naked eye by printing layers from either a regular or an infrared resin using a modified SLA printer. On top of requiring custom materials and hardware, this method also has long processing times because the barcodes projected onto the 3D surface are distorted when captured by a camera, which requires more complex image processing to extract the code.

% However, these techniques require long imaging or decoding times.

Table~\ref{tab:LiteratureComparion} summarizes the capabilities and limitations of the different unobtrusive tags. As can be seen in Table~\ref{tab:LiteratureComparion}, \textit{InfraredTags} is the only unobtrusive approach that allows for simple fabrication of the tags as well as cheap and quick scanning while using existing 2D codes (QR codes, ArUco markers) and facilitating multiple codes to be embedded within the same object. InfraredTags accomplishes this by using infrared imaging, which we explain in the next section.

\begin{figure*}[h]
  \centering
  \includegraphics[width=0.837\textwidth]{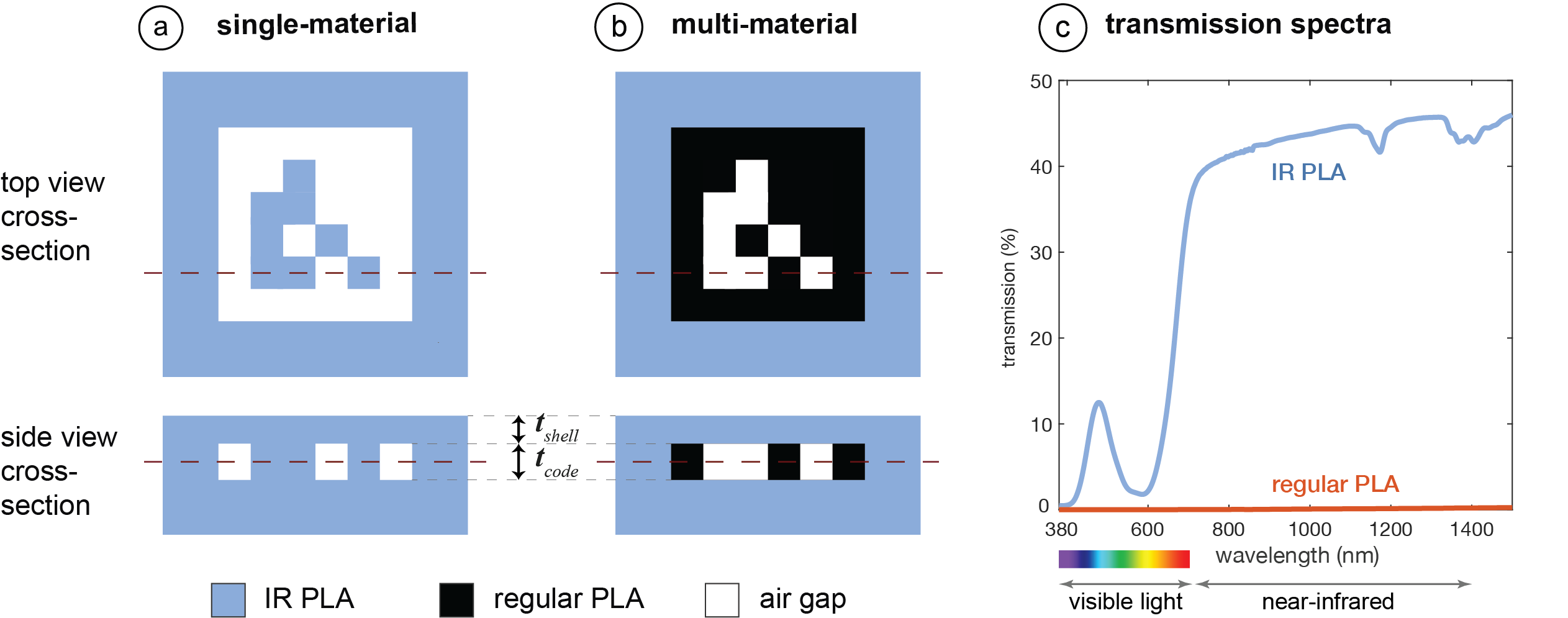}
  \caption{Material composition of the tags for a sample ArUco marker. We modify the interior of the object to embed the tag based on (a) single- or (b) multi-material printing. (c) The transmission spectrum of the IR PLA and regular PLA.}
  \Description{(a) Top and side cross-sections of single-material tag. (b) Top and side cross-sections of multi-material tag. (c) The transmission spectrum of IR PLA shows that it is much more transmitting after 700nm.}
  \label{fig:ModelInterior}
\end{figure*}

\subsection{Near-Infrared Imaging}

Near-infrared (NIR) light is a subregion of the infrared band from the electromagnetic spectrum that is just outside the range of what humans can see.
Today, a wide range of materials are used that enable different use cases with NIR light. For example, retro-reflective materials that reflect NIR can be used to create markers % not visible to humans. 
This is used in common motion tracking hardware tools such as \textit{OptiTrack}\footnote{\url{https://optitrack.com/}} where individual NIR retro-reflective beads can be attached to objects to track them. Because the purpose of these is solely tracking, they do not carry any data.
In \textit{miniStudio}~\cite{kim_ministudio_2016}, Kim et al. project imagery on tangible miniatures by tracking IR reflective stickers attached to them, which were augmented with fiducial markers using screen printing.
Silapasuphakornwong et al.~\cite{silapasuphakornwong_embedding_2020} use  a custom filament that fluoresces under near-infrared illumination to embed patterns inside 3D objects, however, the contrast is not high enough to create 2D tags such as QR codes.
In \textit{HideOut}~\cite{willis_hideout_2013}, Willis et al. create hidden IR markers from IR absorbing ink to project digital imagery on physical paper, however, a spray gun is needed to evenly coat the paper surface. In another project, \textit{SidebySide}~\cite{willis_sidebyside_2011}, they use a custom projector to project NIR markers onto walls in order to enable ad-hoc multi-user portable projector games.
Punpongsanon et al.~\cite{punpongsanon_projection-based_2015} measure deformation of elastic materials by tracking dots painted with IR ink, which are invisible to humans.

There are also materials that let NIR light through but block visible light~\cite{infrared_training_institute_infrared_2019, eplastics_plexiglass_2021}. They appear opaque to humans, but can be used to enclose electronics that transmit NIR light, such as TV remotes. In this project, we leverage this property of NIR-translucent materials to embed invisible codes that carry information inside 3D objects.

\section{InfraredTags}

InfraredTags are embedded such that the objects appear opaque and unmodified under visible light but reveal the tag under near-infrared light. We accomplish this by 3D printing the main geometry of the object using an infrared-transmitting filament, while the tag itself is created by leaving air gaps for the bits. Because the main geometry is semitransparent in the infrared region, the near-infrared camera can see through it and capture the air gaps, i.e., tag, which shows up at a different intensity in the image. %Thus, while the tags can be detected with an off-the-shelf near-infrared camera, they are imperceptible to the naked eye.
We refer to the infrared-transmitting filament as \textit{infrared filament} or \textit{IR filament} in the remainder of the paper.

In the next sections, we describe the properties of the IR filament and the appropriate infrared camera, and then discuss how the IR filament can be used either as a standalone single-material print or together with another filament to create markers inside the object.

\subsection{Infrared Filament}
\label{Infrared_Filament}

We acquired the IR filament from manufacturer 3dk.berlin~\cite{3dkberlin_pla_2021} 
%\footnote{https://3dk.berlin/en/special/115-pla-filament-ir-black.html} 
(ca. \$5.86/100g). It is made out of polylactic acid (PLA), the most common FDM printing filament, and can be used at regular 3D printing extrusion temperatures. To the naked eye, the filament has a slightly translucent black color, however, when 3D printed in multiple layers it looks opaque. 

\vspace{5pt}
\noindent\textbf{IR Translucency:} Since the manufacturer does not provide data on the light transmission characteristics for different wavelengths, we manually measured it using a UV/VIS/NIR spectrophotometer (\textit{PerkinElmer Lambda 1050}). The transmission spectra for both the IR PLA and comparable regular black PLA filament are given in  Figure~\ref{fig:ModelInterior}c. Both spectra are for 1mm thick 3D printed samples.
Because the regular PLA has close to 0\% transmission in both visible and near-infrared regions, it always appears opaque. In contrast, the IR PLA transmits near-infrared at a much higher rate ($\sim$45\%) compared to visible light (0\%-15\%), and thus appears translucent in the IR region and mostly opaque in the visible light region.

%\vspace{-0.13cm}

\subsection{Choosing an Infrared Camera}
%Infrared Camera Choice to Match Filament

To choose the image sensor and filter that can see infrared light and thus can read the tag, we considered the following:

\vspace{5pt}
\noindent\textbf{Filter:} Almost all commercial cameras have an infrared cut-off filter to make camera images look similar to human vision. This filter thus prevents near-infrared light from reaching the image sensor. Since for our purposes, we want to capture the infrared light, we can either buy a camera that has this filter already removed, e.g., the \textit{Raspberry Pi NoIR} camera module, or remove the embedded filter from a camera manually.

\vspace{5pt}
\noindent\textbf{Image Sensor:}  Different cameras' sensors have different sensitivity for different parts of the light spectrum.  %different sensitivity to IR light,assuming their IR cut-off filter has been removed.
To best detect the markers, the sensor should have a high sensitivity in the maximum peak of the material's near-infrared transmission. However, as can be seen in Figure 2c, since the transmission is similar across the entire infrared-region, all cameras that can detect light in the IR region would work for our purposes. For instance, off-the-shelf cheap cameras, such as the \textit{Raspberry Pi NoIR} (\$20), can detect up to 800-850nm in the near-infrared range according to several vendors\footnote{This module has an \textit{Omnivision 5647} sensor. \url{https://www.arducam.com/product/arducam-ov5647-noir-m12x0-5-mount-camera-board-w-raspberry-pi/} \\ \url{https://lilliputdirect.com/pinoir-raspberry-pi-infrared-camera}}. More expensive IR cameras that have sensitivity beyond the near-infrared range, such as \textit{FLIR ONE Pro}\footnote{https://www.flir.com/products/flir-one-pro/}, can detect up to 14,000nm but may cost more than \$400. However, since the infrared transmission does not increase much with higher infrared wavelengths, the low-cost camera is sufficient for our purposes.

\subsection{Composition of the Tags and Materials}

To create InfraredTags, we need to create two geometries with different IR transmission properties that form the object. The different IR transmission properties will cause the two geometries to appear with different intensities in the resulting infrared image. We found that there are two ways to accomplish this. %\stefanie{similar to Infrastructs we have two different printing methods}

% We have to ways to do the printing:
% Multi-material printing vs air gap. Refer to figure. We explain below the pros and cons.

\vspace{5pt}
\noindent\textbf{Single-Material Print (IR PLA):} Our first method uses the IR filament for the main geometry of the object, air gaps for the outside bits of the marker, and IR filament for the inside bits of the marker as shown in Figure~\ref{fig:ModelInterior}a.
The contrast between the bits arises from the fact that the IR light transmission reduces by $\sim$45\% per mm of IR filament (Section~\ref{Infrared_Filament}).
Under IR illumination, the light rays first penetrate the IR filament walls of the 3D printed object and then hit the air gap inside the object or the filled interior area. 
When the object is imaged by an IR camera, the light intensity reduces for each pixel differently depending on whether it is located on an air gap or not.
The rays that go through the air gaps lead to brighter pixels since they penetrate through less material than the other rays.
This intensity difference in the resulting image is sufficient to convert the detected air gaps and filled areas into the original tag.

\vspace{5pt}
\noindent\textbf{Multi-Material Print (IR PLA + Regular PLA):} We explored multi-material 3D printing to further improve the contrast of the marker in the image. This second approach uses IR PLA for the main geometry of the object, regular PLA for the outside bits of the marker, and air gaps for the inside bits of the marker, as shown in Figure~\ref{fig:ModelInterior}b. When the user takes an image, the IR rays penetrate the IR filament walls of the 3D printed object, and then either hit the air gap inside the object or the regular PLA.
The air gaps will appear as brighter pixels since they transmit IR light, whereas the regular PLA filament will appear as darker pixels since it is nearly completely opaque in the IR region (Figure~\ref{fig:ModelInterior}c). This leads to a higher contrast than the previously discussed single-material prints.

We also considered filling the air gaps with IR filament to avoid empty spaces inside the object geometry. However, this requires frequent switches between the two material nozzles for regular PLA and IR filament within short time frames, which can lead to smearing. We therefore kept the air gaps for objects that we printed with the dual-material approach (Figure~\ref{fig:ModelInterior}b).

\vspace{0.2cm}
\noindent
\textbf{Code Geometry}: When embedding the code (i.e., the 2D tag) into the geometry of the object, the code and the geometry surrounding it (i.e., the shell) need to have a certain thickness. 

\vspace{0.2cm}
\noindent
\textit{Shell Thickness}: The shell thickness $t_{shell}$ should be large enough to create sufficient opaqueness so that the user cannot see the code with their eyes, but small enough to ensure detectability of the code with the IR camera.

\begin{figure}[t]
  \centering
  \includegraphics[width=1\columnwidth]{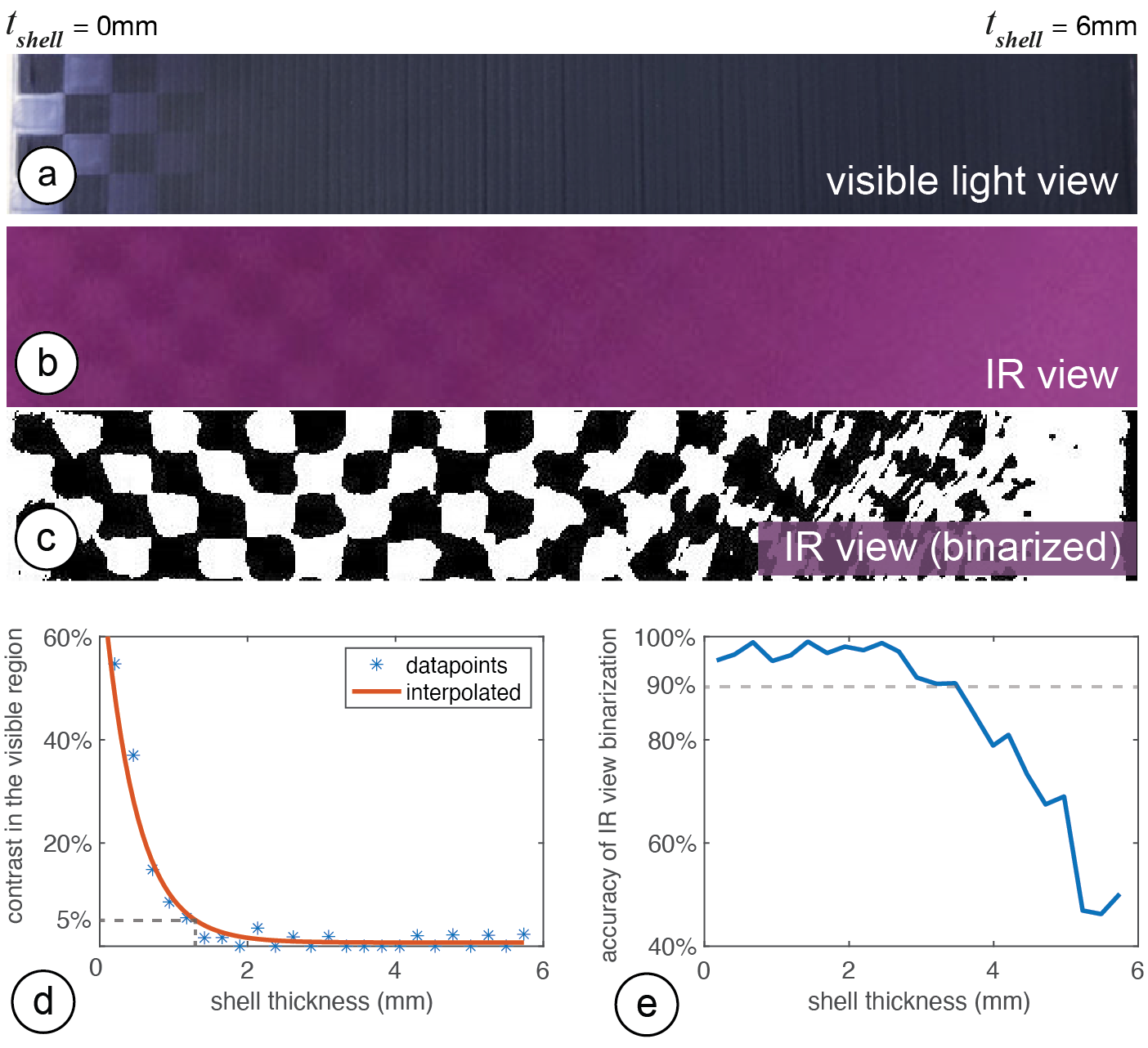}
  %\vspace{-0.1cm}
  \caption{Determining the shell thickness for a multi-material print with white PLA. As $t_{shell}$ increases, the checkerboard pattern becomes less visible in both the (a) visible camera and (b) IR camera image. Thus, it gets more challenging to (d) identify the contrast in the pattern for humans and to (c, e) binarize it correctly from the IR view.}
  \Description{(a) Visible light view: A checkerboard with a black gradient overlay on top that becomes darker toward the right side. (b) IR view: A similar checkerboard pattern, but this time the overlay is violet and it gets more visible toward the right side, but at a slower speed. (c) IR view (binarized): The checkerboard pattern in black in white, but the pattern gets worse and eventually completely white towards the right side. (d) A plot of shell thickness vs. contrast in the visible region; an inverse relationship is shown. (e) A plot of shell thickness vs. accuracy of IR view binarization. It starts at close to 100\% at 0mm shell thickness first, but drops below 90\% and even lower starting around 3.5mm.}
  \label{fig:ShellThicknessValues}
  \vspace{-0.2cm}
\end{figure}

Since the IR filament is slightly translucent to the naked eye, with small $t_{shell}$, it becomes possible for the user to see the code inside the object (Figure~\ref{fig:ShellThicknessValues}a).
Thus, for the lower bound of $t_{shell}$, our goal is to find a value that achieves a contrast in the image smaller than 5\% when the image is taken with a regular camera (i.e., with an IR cut-off filter). The image taken with regular camera represents the visible light region sensitivity, i.e., that of human vision. We chose 5\% because this is the contrast value at which humans cannot differentiate contrast anymore~\cite{bijl_visibility_1989}.

On the opposite side, the larger $t_{shell}$ is, the more IR light it absorbs, and thus the darker the overall image becomes, reducing the contrast of the code in the IR region (Figure~\ref{fig:ShellThicknessValues}b).
Thus, for the upper bound for $t_{shell}$, our goal is to find the value at which the code is no longer detectable in the IR camera image.

To determine these bounds, we 3D printed a checkerboard pattern as an InfraredTag with a shell of varying thickness (range: 0mm-6mm). As shown in Figure~\ref{fig:ShellThicknessValues} for a multi-material print with white PLA, we captured the pattern with both a visible light camera and an IR camera. In  Figure~\ref{fig:ShellThicknessValues}d, we plot the contrast between the "white" and "black" parts of the checkerboard as a function of shell thickness in the visible light camera image. We see that the visible light camera contrast drops to 5\% at approximately 1.32mm thickness, which defines the lower bound, i.e., the minimum thickness needed so that the tag is invisible to humans.
On the other hand, Figure~\ref{fig:ShellThicknessValues}c shows the binarized version of the IR camera image (Figure~\ref{fig:ShellThicknessValues}b).
In  Figure~\ref{fig:ShellThicknessValues}e, we show how the binarization of the checkerboard deteriorates as shell thickness increases. 
The graph shows that a shell thickness of up to 3.5mm could be used to achieve 90\% binarization accuracy, which defines the upper bound. However, for the sake of maximum detectability, we use the lower bound values when fabricating the objects.

For multi-material 3D printing, different filament colors can be used for the regular PLA part (i.e., the code). Each color requires a different shell thickness to prevent users from seeing the code. For instance, because the IR filament appears \textit{black} in the visible light region, it blends more easily with \textit{black} or \textit{blue} PLA, thus requiring a thinner top layer to hide the resulting code than when the code is printed in white PLA. Table~\ref{tab:CodeThickness} shows the minimum shell thickness needed to make codes fabricated from different colors unobtrusive.

\vspace{0.2cm}
\noindent
\textit{Code Thickness}: While the shell thickness affects the overall contrast of the image in the visible region, the code thickness $t_{code}$ determines the contrast between the individual bits of the embedded code in the IR region. 
If the code layer is too thin, there might not be enough contrast between the "white" and "black" bits, and thus the code will not be detectable.

%We conducted a test similar to the one shown in Figure~\ref{fig:ShellThicknessValues}, where we, instead of varying the shell thickness, varied the code thickness to determine which values provide enough contrast.
We conducted a test similar to the one shown in Figure~\ref{fig:ShellThicknessValues} in which we varied the $t_{code}$ instead of $t_{shell}$ to determine which values provide enough contrast.
The values are summarized in Table~\ref{tab:CodeThickness}. Going below the values listed makes the material too thin such that the IR light starts going through the code bits, which reduces the contrast in IR view and thus detectability.
Going above this value is possible but does not improve the contrast further.

\begin{table}[h]
\begin{tabular}{|p{5.5pc}|l|l|}
\cline{2-3}
\multicolumn{1}{c|}{} & {\footnotesize\textbf{Shell thickness} $t_{shell}$} & {\footnotesize\textbf{Code thickness} $t_{code}$} \\ \hline
Single-material\newline {\footnotesize (IR PLA)}            & 1.08 mm & 2.00 mm \\ \hline
Multi-material\newline  {\footnotesize (IR PLA + white PLA)} & 1.32 mm & 0.50 mm \\ \hline
Multi-material\newline  {\footnotesize (IR PLA + black PLA)} & 1.08 mm & 0.50 mm \\ \hline
Multi-material\newline  {\footnotesize (IR PLA + blue PLA)} & 1.20 mm & 0.50 mm \\ \hline
\end{tabular}
\vspace{0.2cm}
\caption{Thickness values for the shell and code layers.}
\label{tab:CodeThickness}
\vspace{-0.5cm}
\end{table}

Lastly, an important observation we made is that IR filament spools ordered from the same manufacturer~\cite{3dkberlin_pla_2021} at different times showed slightly different transmission characteristics. This is likely linked to the possibility that the manufacturer may have adjusted the amount of IR-translucent dye used to make the spools. %and might have later changed the chemical composition.
We suggest that users conduct a similar contrast analysis as shown in Figure~\ref{fig:ShellThicknessValues} to determine the optimal values for each new IR spool.

\section{Embedding and Reading InfraredTags}

We next explain how users can embed codes into 3D objects using our user interface and then discuss our custom add-on for mobile devices and the corresponding image processing pipeline for tag detection.

\subsection{User Interface for Encoding InfraredTags}
\label{User-Interface}

\vspace{5pt}
\noindent\textbf{Import and Position Tags:} The user starts by loading the 3D model (.stl file) into our user interface, which is a custom add-on to an existing 3D editor (\textit{Rhinoceros 3D}). Next, users import the tag as a 2D drawing (.svg) into the editor, which loads the marker into the 3D viewport. The marker is then automatically projected onto the surface of the 3D geometry (Figure 1a). Users can move the code around in the viewport and scale it to place it in the desired location on the 3D object. 

\vspace{5pt}
\noindent\textbf{Select Printing Method:} In the user interface, users can then select the printing method, i.e., if they want to fabricate the object with single material (IR-PLA only) or dual-material printing (IR-PLA + regular PLA). As a result, the user interface generates the geometry to accommodate the selected printing method. For example, for dual-material printing, it generates two .stl files, one for the main geometry and one for the embedded tag. The UI ensures that the tag is accurately spaced from the surface of the object (Table~\ref{tab:CodeThickness}). The user can then slice both files with the 3D printer's slicing software and print the object. 

% To standardize the approach in which we embed 

\begin{figure}[t]
  \centering
  \includegraphics[width=1\columnwidth]{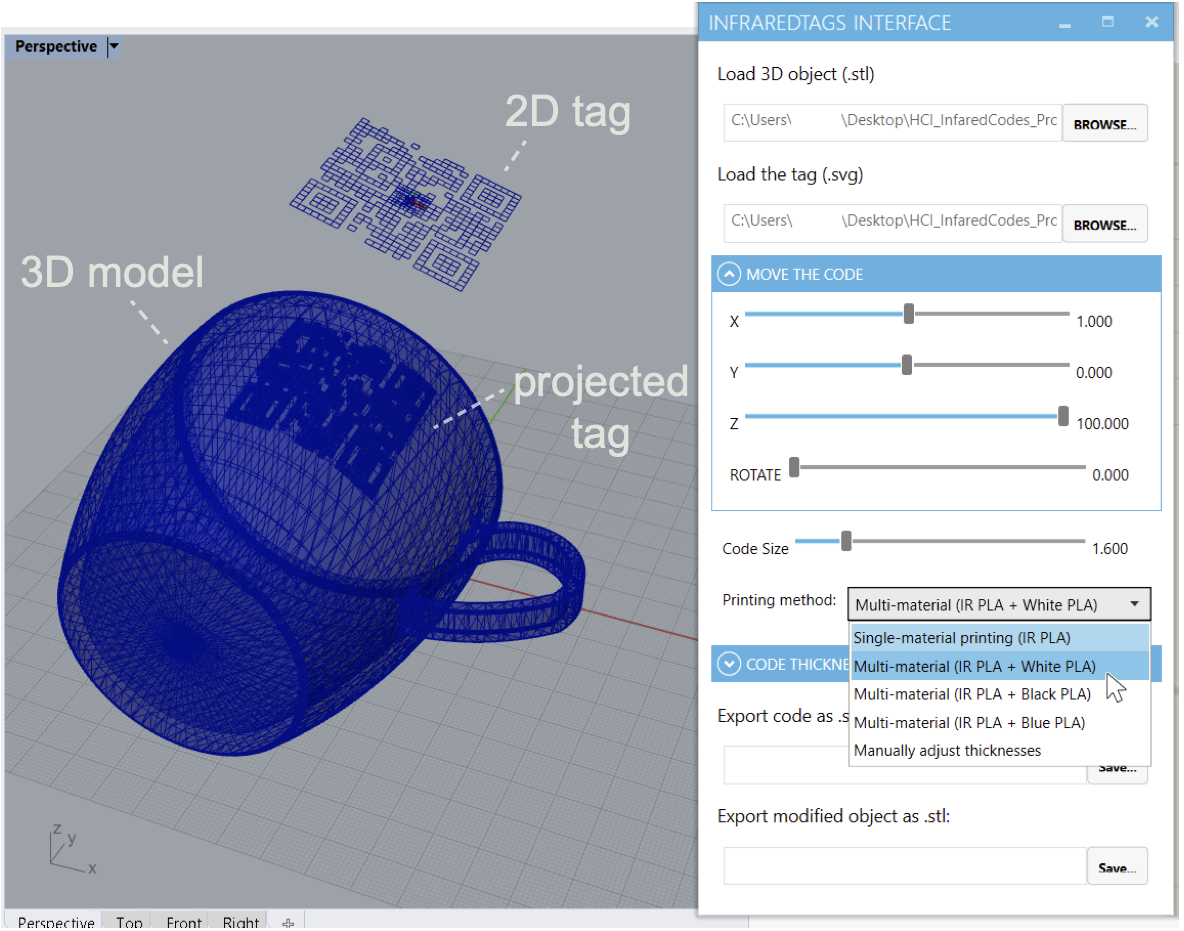}
  \caption{InfraredTags embedding interface.}
  \Description{A software user interface, where the viewport shows a mug’s digital model. The right panel has inputs for the object file, tag file, and choosing the printing method.}
  \label{fig:UI}
\end{figure}

% The user then launches the 3D printer's slicer software and loads the generated models. And starts printing. Ready to be detected without any post-processing.

\subsection{IR Imaging Module for Reading the Tags}
\label{DetectionInterfaceIRModule}
InfraredTags can be read with digital devices that have an infrared camera attached to them. Even conventional USB webcams for personal computers can be used for this purpose by manually removing their infrared cut-off filter\footnote{\url{https://publiclab.org/wiki/webcam-filter-removal}}.

Today, several recent smartphones already come with an IR camera either on the front (\textit{Apple iPhone X}) or the rear (\textit{OnePlus 8 Pro}), however, the phones' APIs may not allow developers to access these for non-native applications. Furthermore, not all mobile phones contain such a camera at the moment. To make our method compatible independent of the platform, we built an additional imaging add-on that can easily be attached to existing mobile phones.

\begin{figure}[t]
  \centering
  \includegraphics[width=1\columnwidth]{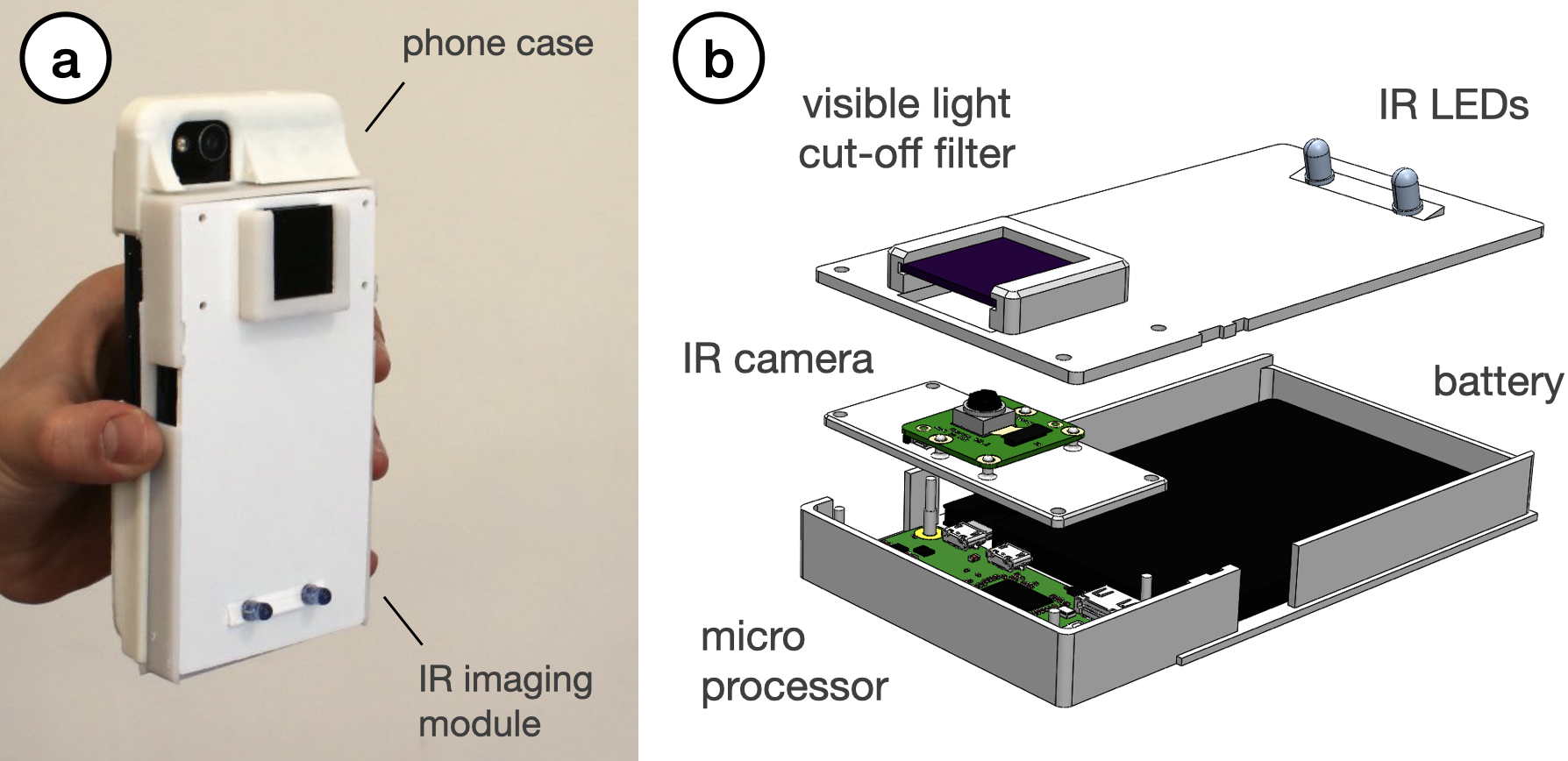}
  \caption{Infrared imaging module. (a) The module is attached onto a flexible case that can be 3D printed based on the user's mobile device. (b) The module's hardware components. }
  \Description{(a) The back of the phone, which shows a white case (IR imaging module is labeled). (b) An exploded view of the IR imaging module’s hardware components (microprocessor, camera, LEDs, battery, filter).}
  \label{fig:IRImagingModule}
\end{figure}

\vspace{5pt}
\noindent\textbf{Attaching the IR camera module:}  As shown in Figure~\ref{fig:IRImagingModule}, our add-on contains an infrared camera (model: \textit{Raspberry Pi NoIR}). This camera can see infrared light since it has the IR cut-off filter removed that normally blocks IR light in regular cameras. Additionally, to remove the noise from visible light and improve detection, we added a visible light cut-off filter\footnote{\url{https://www.edmundoptics.com/p/1quot-x-1quot-optical-cast-plastic-ir-longpass-filter/5421/}}, as well as 2 IR LEDs (940nm) which illuminate the object when it is dark.
This add-on has two 3D printed parts: a smartphone case from flexible TPU filament that can be reprinted based on the user's phone model, and the imaging module from rigid PLA filament that can be slid into this case. The imaging module has a \textit{Raspberry Pi Zero} board and a battery and weighs 132g.

% Any device or computer that has an IR camera connected to is can be used to detect and read InfraredTags.

\vspace{5pt}
\noindent\textbf{Detecting the Tag:} To detect the tag, users open the InfraredTags detection application on their mobile phones and point the camera to their object. The application shows the phone camera's view, which is what the user sees with their eyes instead of the IR view (Figure~\ref{fig:teaser}a). This way, more information can be overlaid on top of the regular view for AR applications.
Under the hood, the imaging module continuously streams the images to our image processing server. If the server detects any tags, it sends the location and the encoded message to the smartphone app to show to the user.

\section{Software Implementation}

In this section, we explain how we implemented the code embedding user interface, as well as our infrared imaging module and image processing pipeline.

\subsection{UI Implementation}

Our embedding user interface is based on 
\textit{Rhinoceros 3D} CAD software\footnote{\url{https://www.rhino3d.com/}} (referred to as \emph{Rhino}) and \emph{Grasshopper}\footnote{\url{https://www.grasshopper3d.com/}} which is a visual programming language that runs within \textit{Rhino}.

\vspace{5pt}
\noindent\textbf{Importing the Tag \& the 3D Model:}
After the user loads an STL file representing the 3D object, our software converts it into a mesh utilizing a Python subprocess function call. The script then centers the mesh along its center of mass. When the user imports a tag as an SVG file, it creates a plane that contains the paths that represent its bits, i.e., the air gaps. While the user is positioning the code, our software always orients the plane of the code to face the mesh's surface. For this, it uses the normal on the mesh that is closest to the plane that holds the code.

% To project the code onto the 3D object's surface, our software uses the closest normal on the mesh to the plane that holds the code. %Similarly, this code, because it is attached to the plane, will also face away from mesh.
% Once the code is properly oriented, we are ready to project it onto the mesh surface. 

\vspace{5pt}
\noindent\textbf{Embedding the Tag into the Object}:
Depending on the type of embedding selected (i.e., single-material or multi-material 3D printing), the tag is projected into the object in one of two ways:

\vspace{0.2cm}
\noindent
\textit{Single-Material:}
Our software first projects the tag onto the curved surface of the mesh and then translates it along the inverted closest mesh normal (i.e., pointing it towards the mesh) by the shell thickness ($t_{shell}$, see Table~\ref{tab:CodeThickness}). We then extrude the tag along the inverted normal by the code thickness ($t_{code}$), which creates a new mesh inside the object representing the air gaps inside the 3D geometry. To subtract the geometry that represents the air gaps from the overall geometry of the 3D object, we first invert the normals of the air gap mesh and then use a Boolean join mesh function to add the holes to the overall object geometry. This results in the completed mesh with the code, i.e., air gaps, embedded that the user can export as a single printable STL file.

\vspace{0.2cm}
\noindent
\textit{Multi-Material:} For multi-material prints, our software generates two meshes as illustrated in Figure~\ref{fig:ModelInterior}b: one for the tag (printed in regular PLA) and one for the shell (printed in IR PLA).
We start by following the steps described for the single-material approach, i.e., project the path representing the tag's bits onto the curved surface, translate it inwards, and extrude it to generate the tag mesh. Next, we find the bounding box of the this mesh, invert its normals, and join it with the main object's mesh. This creates a new mesh (i.e., the IR PLA shell), which once printed will have space inside where the regular PLA tag can sit.

\subsection{Mobile IR Imaging}
The mobile application used for capturing the tags is Web-based and has been developed using \textit{JavaScript}. It uses \textit{Socket.IO}\footnote{https://socket.io/} to communicate with a server that runs the image processing pipeline for tag detection explained in Section~\ref{ImageProcessing}.

The image processing server receives the images from the live stream shared by the microprocessor (\textit{Raspberry Pi Zero W}) on the imaging module and constantly runs the detection algorithm. If a tag is detected, the server sends the tag's location and the decoded message to the Web application, and shows it subsequently to the user. Because the imaging module does not use the resources of the user's personal device and is Web-based, it is platform-independent and can be used with different mobile devices.

\subsection{Image Processing Pipeline}
\label{ImageProcessing}

\begin{figure}[h]
  \centering
  \includegraphics[width=1\columnwidth]{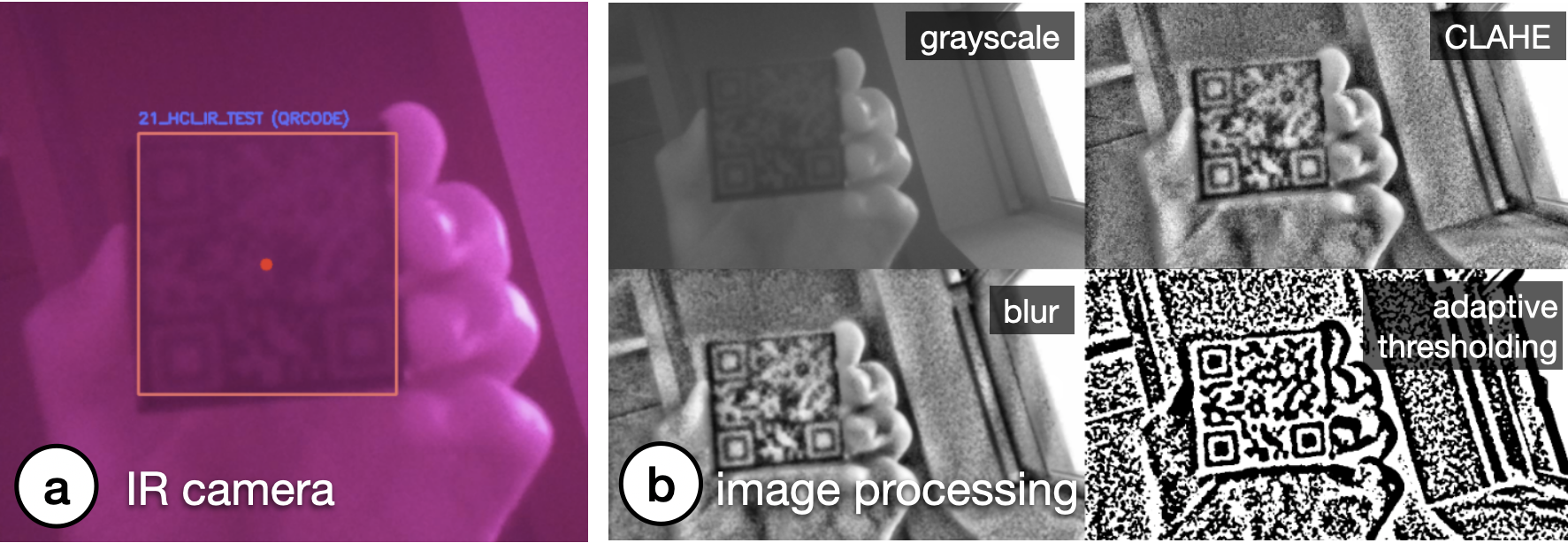}
  \caption{Image processing to read the tags. (a) Infrared camera view. (b)  Individual processing steps needed to decode the QR code message: \textit{"HCI\_IR\_TEST"}.}
  \Description{(a) IR camera view of a QR code. (b) Four image processing steps. (c) Regular camera view, which looks completely black.}
  \label{fig:ImageProcessing}
\end{figure}

InfraredTags are identified from the images captured by the IR camera on the mobile phone or attached imaging module. Although the tags are visible in the captured images, they need further processing to increase the contrast to be robustly read. We use \textit{OpenCV}~\cite{bradski_opencv_2000} to perform these image processing steps as shown in Figure~\ref{fig:ImageProcessing}.

\vspace{0.2cm}
\noindent
\textbf{Pre-processing the Image:} We first convert the image to grayscale and apply a  contrast limited adaptive histogram equalization (CLAHE) filter~\cite{pizer_adaptive_1987} to improve the local contrast (\texttt{clipLimit} = 20, \texttt{tileGridSize} = (8,8)). %To make the bits of the code clearer,
For our pipeline, CLAHE is more appropriate than a standard histogram equalization as it redistributes the pixel intensity values based on distinct sections of the image~\cite{garcia-martin_vein_2020}.
%This takes in two parameters, Clip Limit and Tile Grid Size, to improve the local contrast of the image. Through testing we have found that a Clip Limit of 20 and a Grid Size of (8, 8) consistently produces high contrast for detection.
To reduce the high-frequency noise that arises due to CLAHE, we smooth the image with a Gaussian blur filter. We then binarize the image using Gaussian adaptive thresholding to obtain black-and-white pixels that contain the code (\texttt{constantSubtracted}=4).
%This takes in two parameters, block size and a constant offset. For the constant, we have found that a value of four consistently produces clear binarization and easily readable codes.
% The block size requires more calibration for clear detection, due to camera position and luminosity at time of detection. Typically, however, we have found that values between 21 and 101 create enough clarity for detection.

\vspace{0.2cm}
\noindent
\textbf{Code Extraction:} Once the binary image is generated, it is used to read the respective code using existing libraries, such as \textit{Dynamsoft}\footnote{\url{https://www.dynamsoft.com/barcode-reader/overview/}} or \textit{ZBar}\footnote{\url{http://zbar.sourceforge.net/}}. On average, it takes 6ms to decode a 4x4 ArUco marker and 14ms to decode a 21x21 QR code from a single original frame. The images we use as input are %640x360 
512x288 pixels; in the future, the detection could be made even faster by downsampling the image to a dimension optimal for both readability and speed.

\vspace{0.2cm}
\noindent
\textbf{The Effect of Tag Distance}:
The readability of the binarized tag depends on the parameters used for the pre-processing filters. More specifically, we found that a different Gaussian kernel for the blur (\texttt{ksize}) and block size for the adaptive threshold (\texttt{blockSize}) need to be used depending on the size of the tag in the captured image, i.e., the distance between the tag and the camera.
This is especially important for QR codes since they generally have more and smaller bits that need to be correctly segmented.

% \hl{As described in Section}~\ref{ImageProcessing}, \hl{the choice of \textit{OpenCV} filters parameters affect the quality of the binarized image that is used as input to the QR code or ArUco marker reader.

% Please add the following required packages to your document preamble:
% \usepackage[table,xcdraw]{xcolor}
% If you use beamer only pass "xcolor=table" option, i.e. \documentclass[xcolor=table]{beamer}
\begin{table}[b]
\begin{tabular}{|p{3.7cm}|p{1.5cm}|}
%\begin{tabular}{|l|l|}
\hline
\rowcolor[HTML]{EFEFEF} 
\textbf{Filter combinations}                                                                                        & \textbf{Accuracy} \\ \hline
{\small (\texttt{ksize}=3, \texttt{blockSize}=23)}                                                                                             & 56.45\%            \\ \hline
{\small (\texttt{ksize}=3, \texttt{blockSize}=23),}\newline {\small(\texttt{ksize}=1,} {\small\texttt{blockSize}=37)}                           & 70.97\%            \\ \hline
{\small(\texttt{ksize}=3, \texttt{blockSize}=23),}\newline {\small(\texttt{ksize}=1, \texttt{blockSize}=37),}\newline {\small(\texttt{ksize}=3, \texttt{blockSize}=21)} & 79.03\%            \\ \hline
\end{tabular}
\vspace{0.2cm}
\caption{Filter combinations and QR code detection accuracy}
\vspace{-0.57cm}
\label{tab:filter_combinations}
\end{table}

One strategy to increase detection accuracy is to iterate through different combinations of the filter parameters. To identify the effect of the number of filter parameter combinations on detection accuracy, we ran the following experiment: We captured 124 images of a 21x21 QR code from different distances (15-80cm from the camera). We then generated 200 different filter parameter combinations and used them separately to process the captured images. We then evaluated which filter parameters correctly binarized the QR code. We found that even with a small number of filter combinations, we can have sufficient detection results comparable to existing QR code detection algorithms. For instance, three different filter combinations (Table~\ref{tab:filter_combinations}) achieve an accuracy up to 79.03\% (existing QR code readers achieve <57\% for blurred tags\footnote{Peter Abeles. 2019. Study of QR Code Scanning Performance in Different Environments. V3. \url{https://boofcv.org/index.php?title=Performance:QrCode}}). It is possible to further increase the number of filter parameter combinations to improve the accuracy further at the expense of detection time.
% Please note that the other filters' parameters were held constant as given in Section~\ref{ImageProcessing}.

\begin{figure*}[t]
  \centering
  \includegraphics[width=0.97\textwidth]{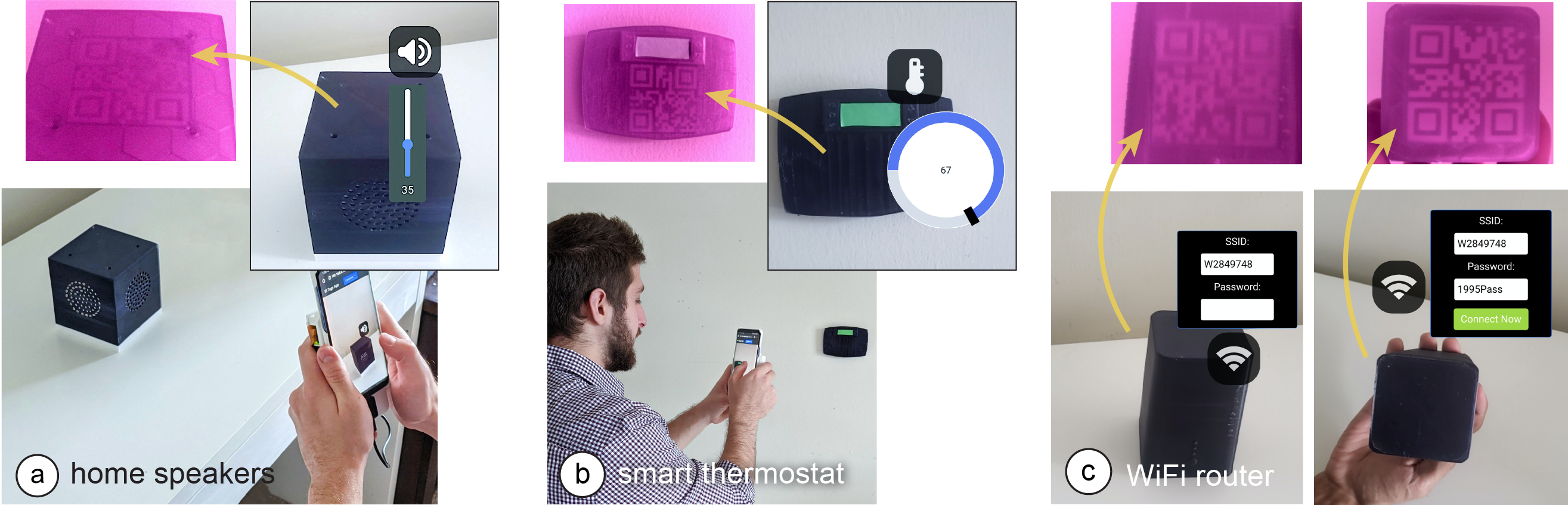}
  \caption{Controlling appliances using a mobile AR application. The user points at (a) the home  speakers to adjust its volume, and the (b) thermostat to adjust the temperature. The infrared camera in the phone's case identifies the appliances by reading the embedded QR codes. (c) Pairing a phone with a WiFi router, whose SSID is visible from all sides but the password is visible only from its bottom.}
  \Description{A user adjusting the devices’ controls. (a) Home speakers. (b) Smart thermostat. (c) WiFi router.}
  \label{fig:AppliancesAR}
\end{figure*}

\section{Applications}
We demonstrate how InfraredTags enable different use cases for interactions with objects and devices, storing data in them, and tracking them for sensing user input.

\subsection{Distant Augmented Reality (AR) Interactions with Physical Devices}
\label{Applications_Appliances}

InfraredTags can be embedded into physical devices and appliances to identify them individually through the embedded unique IDs and show the corresponding device controls that can be directly manipulated by the user.

In the application shown in Figure~\ref{fig:AppliancesAR}a and b, a user points their smartphone camera at the room and smart home appliances are identified through their InfraredTags, which are imperceptible to the human eye. A control menu is shown in the AR view, where the user can adjust the volume of the speaker or set a temperature for the thermostat.
InfraredTags could also allow multiple appliances of the same model (e.g., multiple smart speakers or lamps) in the room to be identified individually, which is not possible with standard computer vision-based object classification approaches.

\vspace{0.2cm}
\textit{Multiple tags on a single object for spatial awareness:} 
Furthermore, InfraredTags enable \textit{multiple tags} to be embedded in the same object. This enables different applications. For instance, when an object is partially occluded, multiple tags in the object can allow the capture of tags from different angles. Another application is to enable spatially aware AR controls where different settings appear at different locations within the same object. For example, Figure~\ref{fig:AppliancesAR}c illustrates how the front, side, and top faces of a WiFi router only have its network name (SSID) information, whereas its bottom also shows the password information, which can automatically pair the router to the phone. This enables quick pairing and authentication with devices without users having to type out complex character strings, while maintaining the physical use metaphors, such as the paper slip containing the password typically attached to the base of the router. 
%\hl{Figure XX shows another spatially-aware application where pointing the camera at the bottom of the lamp brings up on/off and brightness controls whereas pointing it at the top brings up color change controls enabling quick shortcuts that again resemble physical use.}
% like a short cut "quick control"
While we demonstrate this application for mobile AR, InfraredTags could also enable lower friction, distant interactions with physical devices for head-mounted AR glasses.

\subsection{Embedding Metadata about Objects}
\label{Applications-Metadata}
\begin{figure}[b]
  \centering
  \includegraphics[width=0.95\columnwidth]{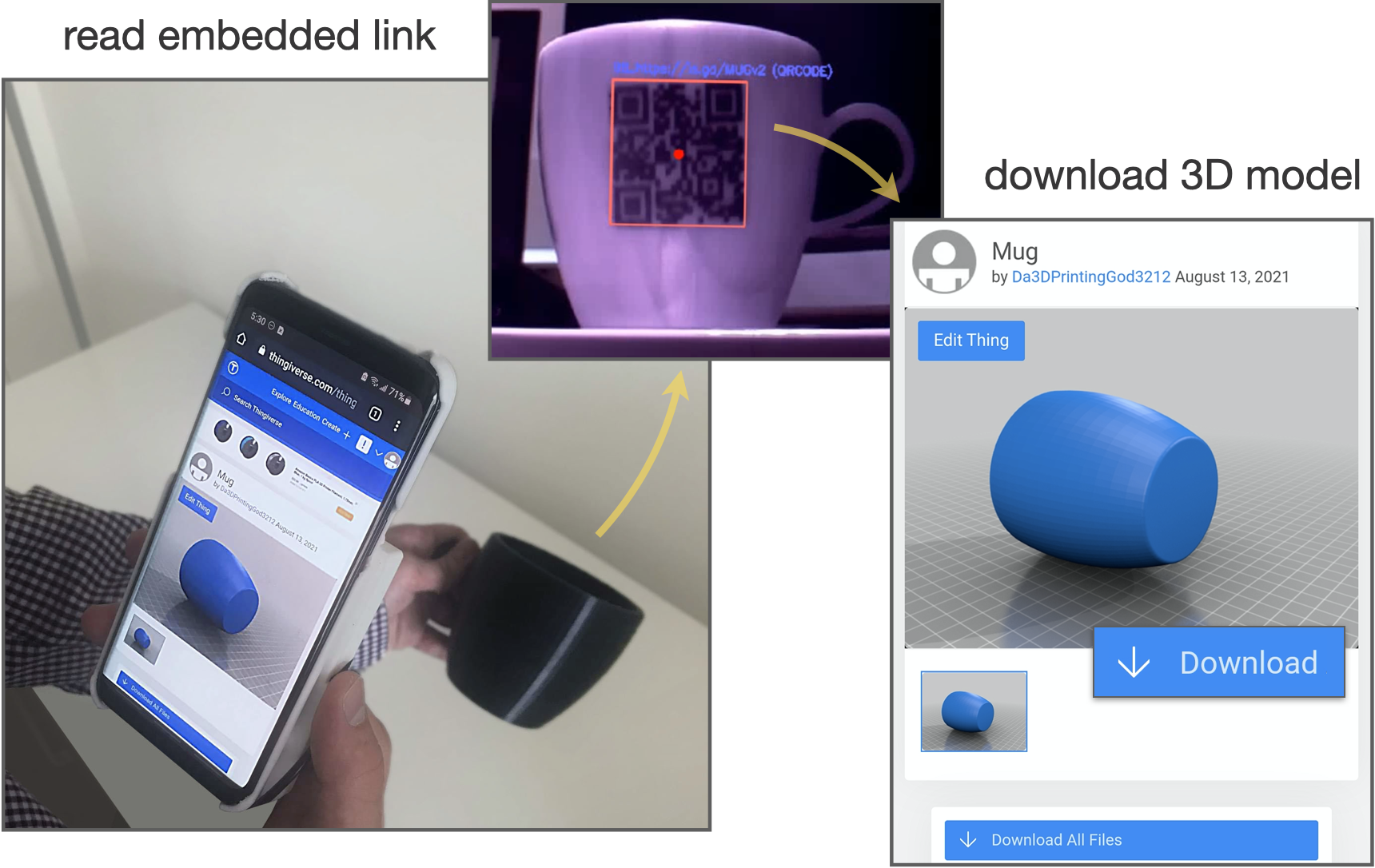}
  \caption{Embedded metadata about the object itself: The user is redirected to the \textit{Thingiverse} model that was used to fabricate the object.}
  \Description{The user is pointing the phone at the mug, which launches a Thingiverse 3D model download page.}
  \label{fig:FabricationURLMug}
  %\vspace{-0.2cm}
\end{figure}

Spatially embedding metadata or documentation information within the object itself can provide richer contextualization and allow information sharing~\cite{ettehadi_documented_2021}. For example, we can embed the object's fabrication/origin link (e.g., a shortened \textit{Thingiverse} URL) as an InfraredTag  for users to look up in case they would like to get more information from its creator or 3D print it themselves as shown in
Figure~\ref{fig:FabricationURLMug}.
Other types of metadata that could be embedded include user manuals, expiry dates, date of fabrication, materials used to fabricate the object, weight, or size information.

\subsection{Tangible Interactions: Use Anything as a Game Controller}
\label{Applications-GameController}

Because fiducial markers can be embedded as InfraredTags, they can be used to track the object's movement. Thus, any passive object can be used as a controller that can be held by users when playing video games.

\begin{figure}[b]
  \centering
  \includegraphics[width=1\columnwidth]{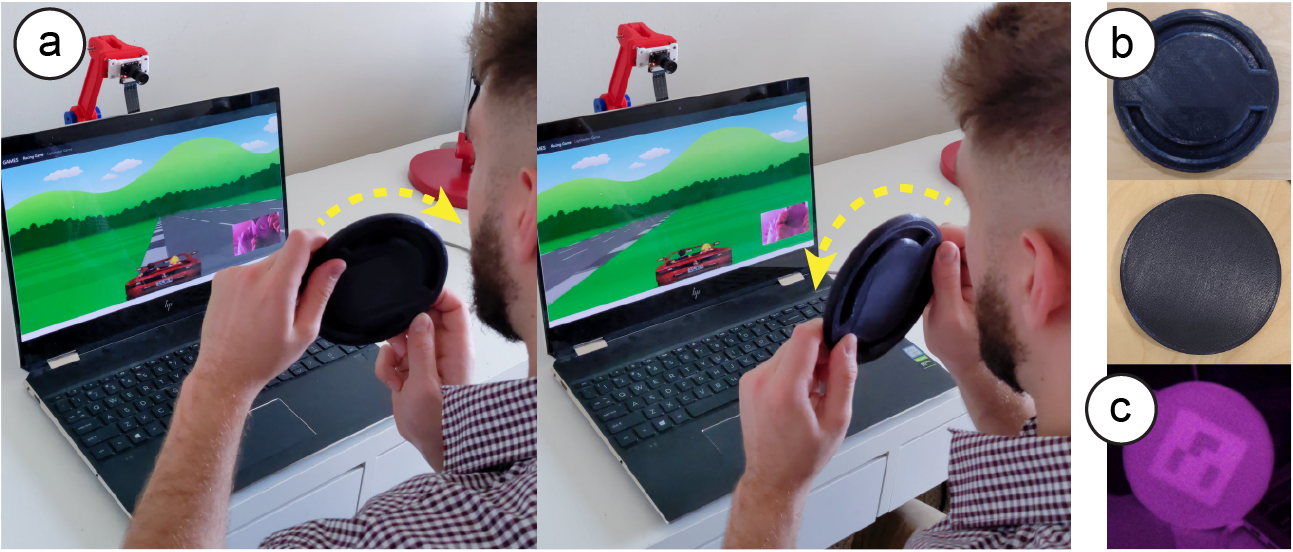}
  \caption{Using passive objects (a) as  a game controller. (b)~This wheel is black under visible light and has no electronic components. (c)~The fiducial marker embedded inside is only visible to an infrared camera.}
  \Description{(a) A man rotating a wheel. (b) A black circular plate. (c) An ArUco marker is visible.}
  \label{fig:CarGame}
\end{figure}

Figure~\ref{fig:CarGame} shows a 3D printed wheel with no electronics, being used as a game controller. The wheel contains an ArUco marker InfraredTag which is used to track the wheel's location and orientation. Even though the wheel is rotationally symmetric, the infrared camera can see the square marker inside and infer the wheel's position and orientation. Our method does not require any electronics as opposed to other approaches~\cite{yamaoka_foldtronics_2019}.

While we demonstrate an application where the user faces a screen with a camera behind it, this could be used to enable passive objects to serve as controllers for AR/VR headsets with egocentric cameras.
Such an application scenario could be particularly suitable for headsets like \textit{HoloLens 2}, which comes with an integrated infrared camera~\cite{ungureanu_hololens_2020} that could be utilized for InfraredTag detection in the future. Even though the tag would be facing the user, it would not be visible to the user but can still be identified by the headset.

\section{Evaluation of the Detection}
\label{TechnicalEvaluation}

In this section, we evaluate how InfraredTag detection is affected by fabrication- and environment-related factors.

\vspace{0.2cm}
\textbf{Marker size}: By following a test procedure similar to the one shown in Figure~\ref{fig:ShellThicknessValues}, we determined that the smallest detectable 4x4 ArUco marker printable is 9mm wide for single-material prints and 6mm wide for multi-material prints.
The resolution for multi-material prints is better than single-material ones because the large transmission difference between the two distinct materials makes it easier for the image sensor to resolve the border between the marker bits.
On the other hand, in single-material prints, the luminosity of an air gap resembles a 2D Gaussian distribution, i.e., the intensity gets higher towards the center. Thus, larger bits are needed to discern the borders between a single-material marker's bits.

\begin{figure}[t]
  \centering
  \includegraphics[width=1\columnwidth]{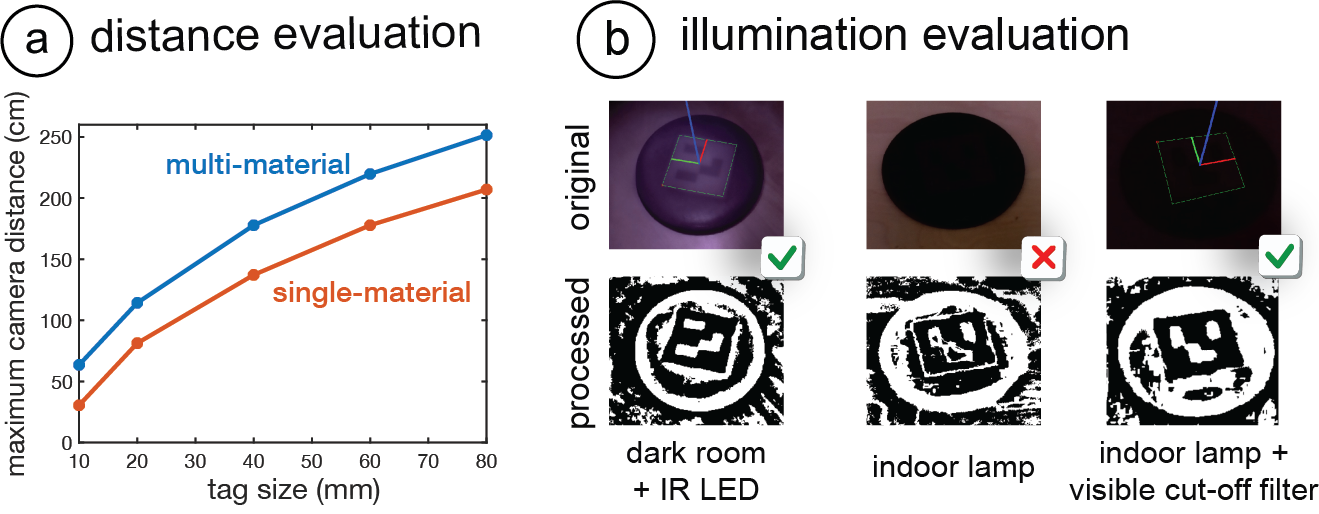}
  \caption{Detection evaluation. (a) Maximum detection distance for single- and multi-material \textit{ArUco} markers. (b) Cases where the IR LED and visible cut-off filter improve detection.}
  \Description{(a) The plot shows that larger codes can be detected from a longer distance. Distances are larger for multi-material than single-material prints. (b) Three cases of illumination: dark room with IR LED, indoor lamp, indoor limp with visible cut-off filter.}
  \label{fig:DetectionEvaluation}
\end{figure}

\vspace{0.2cm}
\textbf{Distance}: To test the limits of our detection method, we measured the maximum distance tags of different sizes can be detected.
This was done for both single-material (IR PLA) and multi-material (IR PLA + regular black PLA) prints.
The marker size range we evaluated was 10-80mm for 4x4 ArUco markers, which would translate to a range for 42-336mm for 21x21 QR codes (can store up to 25 numeric characters).
The results are given in Figure~\ref{fig:DetectionEvaluation}a, which shows that multi-material codes can be detected from further away than single single-material ones. The results are given for the filter parameters with the best detection outcome (Section~\ref{ImageProcessing}). 

% https://www.qrcode.com/en/about/version.html

\vspace{0.2cm}
\textbf{Lighting conditions:} 
For InfraredTags to be discernible in NIR camera images, there has to be enough NIR illumination in the scene. We measured the minimum NIR intensity needed to detect 4x4 ArUco markers using a lux meter which had a visible light cut-off filter (720nm) attached.
We found that just a tiny amount of NIR is sufficient for this, i.e., that at least 1.1 lux is needed for single-material prints, and 0.2 lux for multi-material prints.

Because sunlight also contains NIR wavelengths, the tags are detectable outdoors and also in an indoor areas that have windows during daytime.
We also noticed that many lamps used for indoor lighting emit enough NIR to detect the codes at nighttime (e.g., 1.5 lux in our office).
Furthermore, the IR LEDs on our imaging module (Section~\ref{DetectionInterfaceIRModule}) provide high enough intensity to sufficiently illuminate multi-material markers even in complete darkness (Figure~\ref{fig:DetectionEvaluation}b). In the future, brighter LEDs can be added to support single-material prints in such difficult detection scenarios. 

The visible light-cut off filter used on our IR imaging module also improves detection in spite of challenging lighting conditions. For instance, the last two columns in Figure~\ref{fig:DetectionEvaluation}b shows how certain print artifacts on the object's surface might create noise in the IR camera image, which is reduced when the cut-off filter is added. This is particularly helpful for single-material prints, which are more challenging to identify.

\section{Discussion}
In this section, we discuss the limitations of our approach and how it could be further developed in future research.

\vspace{5pt}
\noindent\textbf{Print Resolution:}
In this project, we used FDM printers, whose printing resolution is restricted by the size of its nozzle that extrudes the material, and a low-cost camera that has an 8MP resolution.
In the future, even smaller InfraredTags can be fabricated by applying our method to printing technologies with higher resolution, such as stereolithography (SLA).
Correspondingly, higher-resolution cameras with better aperture can be used to identify these smaller details (e.g., Samsung’s latest 200MP smartphone camera sensor~\cite{samsung_isocell_2021}).
%\footnote{ISOCELL HP1 \url{https://www.samsung.com/semiconductor/minisite/isocell/mobile-image-sensors/isocell-hp1/}}).
This would allow embedding more information in the same area.

\vspace{5pt}
\noindent\textbf{Discoverability vs. Unobtrusiveness:}
For InfraredTags to be detected, the user should orient the near-infrared camera such that the embedded marker is in the frame.
However, similar to related projects such as \textit{AirCode}~\cite{li_aircode_2017} and \textit{InfraStructs}~\cite{willis_infrastructs_2013}, this might be challenging since the marker is invisible to users and thus they might not know where exactly on the object to point the camera at.
%Ideally, the user can just aim at any arbitrary location on the object, and the near-infrared camera can still see the tag.
For objects with flat surfaces, this can be compensated for by embedding a marker on each face (e.g., on the six faces of a cube). This way, at least one marker will always be visible to the camera. Similar to how a QR code printed on a sheet of paper is detectable from different angles, the flat InfraredTag will maintain its shape when viewed from different angles (e.g., the router Section~\ref{Applications_Appliances}c).

However, detection of codes on curved objects poses a bigger challenge. This is because a 2D code projected onto a curved surface (e.g., the mug in Section~\ref{Applications-Metadata}) has a warped outline when viewed from an angle far away from its center. 
%As a solution, multiple codes can be added on different areas of the object so that a detectable marker is always visible to the IR camera, as shown in the WiFi router application (Section~\ref{Applications_Appliances}).
As a solution, we plan to pad the whole object surface with the same code, similar to \textit{ChArUco} (a combination of ArUcos and chessboard patterns)~\cite{hu_deep_2019}, so that at least one of the codes appears undistorted in the captured image.
Also, for curved objects, other tag types that are more robust to deformations could be used~\cite{yaldiz_deepformabletag_2021} in the future.
Alternatively, a small visible or tactile marker in the form of a notch could be added to the surface of the object (corresponding to where the code is embedded) to help guide the user to the marker.

\vspace{5pt}
\noindent\textbf{Other Color and Materials:}
While we only used black IR PLA in this project, manufacturers could produce filaments of other colors that have similar transmission characteristics to create more customized or multi-material prints in rigid and flexible forms~\cite{forman_defextiles_2020}.
We also plan to combine the IR PLA filament with IR retro-reflective printing filaments to increase the marker contrast even more.

\section{Conclusion}
In this paper, we presented InfraredTags, a low-cost method to integrate commonly used 2D tags into 3D objects by using infrared transmitting filaments. We explained how this filament can be used by adding air gaps inside the object or by combining it with regular, opaque filaments, which increases the tag contrast even more. We discussed what kind of camera is appropriate for detecting InfraredTags and what kind of code geometry should be used for best detection, while ensuring unobtrusiveness to the naked eye. After introducing our tag embedding user interface and mobile infrared imaging module, we presented a wide range of applications for identifying devices and interacting with them in AR, storing information in physical objects, and tracking them for interactive, tangible games. Finally, we evaluated our method in terms of marker size, detection distance, and lighting conditions.

% \input{09-implementation}
% \input{11-discussion}
% \input{12-conclusion}

% \section{ACKNOWLEDGEMENTS}
% <Removed for blind review.>

%%
%% The acknowledgments section is defined using the "acks" environment
%% (and NOT an unnumbered section). This ensures the proper
%% identification of the section in the article metadata, and the
%% consistent spelling of the heading.
\begin{acks}
    We would like to thank Raul Garcia-Martin for offering insights into image processing approaches for near-infrared imaging and Mackenzie Leake for proofreading our manuscript. We also thank the anonymous authors for their constructive feedback. This work was supported by an Alfred P. Sloan Foundation Research Fellowship.
\end{acks}

%%
%% The next two lines define the bibliography style to be used, and
%% the bibliography file.
\bibliographystyle{ACM-Reference-Format}
\bibliography{InfraredTags}

%\bibliography{lasercutter-speckle,MyLibrary}
%\bibliography{sample-base}

% \appendix
% \input{11-appendix}

\end{document}